# Multi-State Neurons

Robert Worden

Active Inference Institute, Crescent City, CA, USA

rpworden@me.com

Draft 0.5; December 2025

Abstract:

Neurons, as eukaryotic cells, have powerful internal computation capabilities. One neuron can have many distinct states, and brains can use this capability. Processes of neuron growth and maintenance use chemical signalling between cell bodies and synapses, ferrying chemical messengers over microtubules and actin fibres within cells. These processes are computations which, while slower than neural electrical signalling, could allow any neuron to change its state over intervals of seconds or minutes.

Based on its state, a single neuron can selectively de-activate some of its synapses, sculpting a dynamic neural net from the static neural connections of the brain. Without this dynamic selection, the static neural networks in brains are too amorphous and dilute to do the computations of neural cognitive models.

The use of multi-state neurons in animal brains is illustrated in hierarchical Bayesian object recognition. Multi-state neurons may support a design which is more efficient than two-state neurons, and scales better as object complexity increases. Brains could have evolved to use multi-state neurons.

Multi-state neurons could be used in artificial neural networks, to use a kind of non-Hebbian learning which is faster and more focused and controllable than traditional neural net learning. This possibility has not yet been explored in computational models.

.

## 1. Introduction

The eukaryotic cell is an amazingly capable device. Most of the biosphere is built on the capabilities of eukaryotic cells. A large part of an animal's complete genotype may be expressed in a single cell such as a neuron. Yet theoretical neuroscience has largely ignored these capabilities. Why is this?

Almost all computational models of the brain assume a simple model of the neuron – the McCulloch-Pitts neuron [McCulloch & Pitts 1943] with Hebbian learning [Hebb 1949] – a model which is now more than 70 years old. In this model, every neuron acts as a weighted adding machine, computing a weighted sum of all its inputs, to determine whether or not to fire. It acts as a two-state machine, with states 'on' and 'off'. This functionality is so simple that it might almost be done[1] by a probiotic cell, like a specialized bacterium.

It would be surprising if in the design of the brain, evolution should have ignored nearly all the prodigious capabilities of eukaryotic neurons. Those capabilities have existed for more than a billion years. In that time, brains have been required to perform a huge variety of tasks, under tight constraints of size and power consumption. It would seem strange if in response to this vast range of selection pressures, and given the range of eukaryotic cell capabilities, the random search of evolution should never have used the wider eukaryotic cell capabilities available to it, but has always stayed on the narrow path of the McCulloch-Pitts neuron.

The alternative is that evolution has not stayed on a narrow McCulloch-Pitts-Hebbian path; that neurons in the brain have greater capabilities; but that they have not yet been discovered, because we have not looked for them. This paper explores the alternative.

The paper asks three questions.

1. What are the logical computational capabilities required of brains at Marr's [1982] level 2, and can they be met by static neural nets?
2. Is the quantitative neural connectivity in brains efficiently used by McCulloch-Pitts neurons?
3. Is there a better way to meet the brain's computing requirements, using dynamically configurable neural networks? Can the eukaryotic cell capabilities of neurons be used to build dynamic neural networks?

The paper proposes that:

- Chemical messengers are transported over the neural cytoskeleton, making each neuron individually configurable, altering its functionality over short timescales (longer than neural firing timescales, but within seconds or minutes).
- Each neuron may have a large number of states, which selectively de-activate synapses to alter its connectivity and function over short timescales.
- The brain can be seen not as a few large static loosely connected neural networks, but as a number of smaller tightly connected dynamic networks.
- These small networks represent dynamic data structures that are required for many purposes.

This proposal is illustrated in hierarchical Bayesian object recognition [Kersten & Yuille 2003; Friston 2010]. When analysed at Marr's [1982] level 2 of abstract data structures and algorithms, object recognition requires the building and comparison of dynamic multi-level tree structures, representing objects and their parts. How are these tree structures represented by neurons?

I show that if neurons are two-state on-off machines, with tree structures represented by static neural connectivity, the number of neurons required and the number of inter-region axons both grow rapidly with the size of the trees. Using multi-state neurons, neural firing can convey dynamic metadata defining the tree structures, as well as data on the nodes of trees. The use of dynamic data structures allows the trees to be represented and communicated with much better scaling. The required number of neurons grows only linearly with the total number of nodes in all trees. This is more efficient, so it would have evolved if it is possible.

The proposals of this paper are not supported by new experimental evidence. They aim to stimulate researchers to think outside the box of the McCulloch-Pitts neuron, and to build alternative models of the brain.

---

[1] (apart from cell maintenance, using mitochondria and the cytoskeleton)

Multi-state neurons could be used in artificial neural nets, to give a kind of non-Hebbian learning which is faster and more focused than traditional neural net learning. This possibility has not yet been explored in computational models.

## 2. Capabilities of the Eukaryotic Cell

Eukaryotic cells do everything in the body. They detect light, they produce physical forces, they carry out thousands of chemical reactions, and they convey electric signals. It is possible to imagine (if only for a moment) that each single eukaryotic cell does only a very limited function, expressing only a small part of an animal's total genotype.

A little thought shows that this is not the case. Every eukaryotic cell has considerable computational capability. It carries out complex computations to reach a state where it can perform its function, to maintain its capability to function, and to know when to stop working and die. These computations are seen most clearly in growth and morphogenesis.

In a complex animal, any cell on its own would be useless. It functions in an environment of other cells, of its own cell type and of other cell types. It senses the surrounding cells through its membranes – and based on these signals, it makes complex decisions about how to grow and function. These are computations. They are done by chemical reactions, catalysed by enzymes. This is not just a matter of a few simple reactions happening in a chemical soup. The cell has a complex internal structure of organelles and cytoskeleton. In this structure, information is created in some places, and transported to other places [Puram & Bonni 2013]. The cytoskeleton is constantly ferrying information in the form of chemical messengers between organelles in the cell. Even the internal delivery system must make computations, to deliver chemical messengers to the right places.

This complex intracellular computation creates the physical structure and shape of the cell, and its relationships to surrounding cells, building them together to make a complex functioning machine. The result of computation is a large amount of information, expressed as the shape of the body organs and their functions. We still do not understand the wonder of morphogenesis [Pulam & Sobel 2010]. It is evidently a prodigious computational feat. Modern engineering can create complex machines – but cannot make them do all the computations needed to self-assemble from undifferentiated cells, like assembling an airliner in mid-air.

This is particularly impressive in the morphogenesis of the brain [Poulain & Sobel 2010; Puran & Bonni 2013; Ribatti & Guidolin 2022] – where neurons of thousands of different types all grow with precise geometry over long distances, seek out specific other types of neuron, and selectively make synapses with them. The information content of the form of a brain is very large, and it is the result of a powerful computation.

Yet the prevailing paradigm of neuroscience is that once a brain has grown, all that chemical computation capability is used only for cell maintenance – and that the only dynamic computation done by a neuron is a crude McCulloch-Pitts summation of its inputs.

There could be only one reason for such a perverse-seeming neglect of useful computing function. The reason would be that only Hodgkin-Huxley electric signalling is fast enough [Hodgkin & Huxley 1952], and that chemical computation in neurons is too slow to be useful. That reason does not stand up quantitatively. While inter-neuron electrical computation can be done in milliseconds, chemical computation inside neurons can shape the electrical computation over timescales of seconds or minutes– which is a useful timescale for cognition.

## 3. Multi-State Neurons: How They Could Work

This section contains a proposal for how a multi-state neuron could work. It is shown in figure 1.

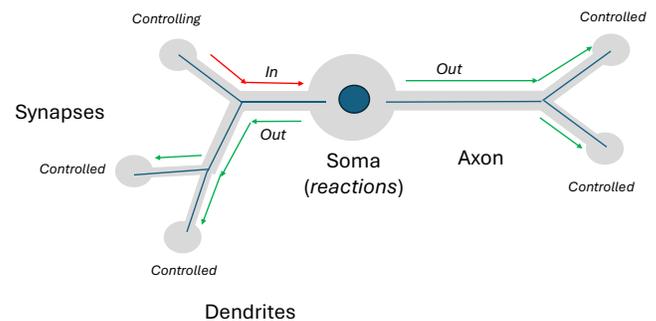

*Figure 1: How a multi-state neuron could work.*

The neuron cell body is a large gray circle in the figure, and the synapses are small gray circles. The lines in the gray tubes are microtubules and actin fibres, which convey chemical messengers to and from the cell body.

The synapses are divided into **controlling** (input) synapses and **controlled** synapses. When a spike from another neuron arrives at a controlling synapse, that synapse emits a chemical messenger or messengers which is transported to the cell body (red arrows). Near the cell body, chemical reactions take place, catalysed by enzymes from expressed genes. These reactions amplify and convert the chemical signals from the controlling synapses.

The results of the central chemical reactions are transported outwards (green arrows) towards the controlled synapses, which are mainly on the dendrites, but which could also be on the axons.

Based on a combination of chemical messengers which arrive at a controlled synapse, it may be rendered inactive.

The result is that information coming into the controlling synapses switches the neuron into one of a number of states, defined by chemical messengers which selectively deactivate the controlled synapses.

There may be a large number of neuron states, and many different types of controlled synapses, which are differentially gated depending on the state. Approximately, if there are N different chemical messengers, that can define as many as $2^N$ different states of the neuron, and as many as $2^N$ different types of controlled synapse, whose action depends on the state. Similarly, controlling synapses may be of different types, depending on the chemical messengers they emit. In this way, the types of neuron which are distinguished anatomically may be divided into a larger number of subtypes.

At any moment, of all the synapses observed anatomically in the brain, only a small fraction may be active - 'carving out' working neural circuits from a possibly amorphous background connectivity. This is shown schematically in figure 2.

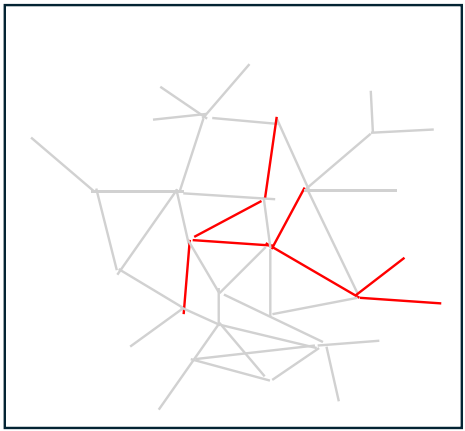

*Figure 2: At any moment, some sparse dynamic neural nets (red lines) are carved out by multi-state neurons, from the static background net of the brain (gray lines). Only a small proportion of synapses are active.*

How fast can a neuron change state? If the changes only involve synapses on dendrites, we can ignore the times needed for chemical messengers to travel over long axons, and we may ignore the times for the enzyme-catalysed reactions to take place. State switching times depend mainly on the messenger transport times in dendrites. For typical neurons in a mammalian brain, dendrite lengths are of the order of 500 μm from the cell soma, and transport speeds can be as large as 10 μm/second. This means that state switching times may be of the order of seconds or minutes. These times may not be useful for immediate responses to sensory stimuli; but for memory and learning, they may be useful.

Multi-state computing capability could arise not only within neurons, but also from their interactions with glial cells, notably astrocytes. In the past, astrocytes were thought mainly to serve neuron maintenance, metabolism and support, but this view is changing as more is learned about them [Gradisnik & Velnar 2023]. Astrocytes may be as many as 40% of the cells in the human brain [Avezedo et al 2009], and they have, as their name implies, complex star-like shapes, making contact with large numbers of neurons. Some astrocytes in the human brain contact as many as a million synapses – which means that they interact with thousands of neurons.

Astrocytes are closely linked to synapses; they could play a role not just in synapse growth and maintenance, but also in controlling inter-neuron synaptic communication. If astrocytes served only some maintenance and support function, like a local plumber, there would be no need for them to have complex star-like shapes spanning large distances. Their complex ramifying shape may convey complex information between neurons across these distances, using chemical messengers conveyed over the cytoskeletons of the astrocytes.

Astrocytes could have chemical states controlled by synapses on neighboring neurons, and could control synapses on other neurons. The firing of a controlling synapse creates chemical messengers which are ferried towards the astrocyte nucleus. Near the nucleus, chemical reactions create other messengers which are transported to controlled synapses, which then gate the communication between neurons.

The combination of multi-state neurons and multi-state astrocytes could give powerful computation capabilities. The chemical reactions might compute not just in a digital on/off manner, but by graded changes in the concentrations of reactants.

## 4. Requirements for Neural Computing

David Marr [1982] proposed that computation in the brain can be studied on three levels:

1. What Marr called the **computational** level – which might now be called the level of **computational requirement** – what the brain is required to do, to ensure the best possible fitness of its owner.
2. The level of **algorithms and data structures** – which are stated abstractly, independent of any specific implementation, and which meet the requirements of level 1.
3. The level of **neural implementation** – how neurons carry out the abstract computations defined at level 2.

This three-level framework is analogous to the many levels of 'virtual machine' used in modern computing, and has been a very successful way to study the brain.

One current form of it is the theory of Bayesian cognition. It can be shown at Marr's level 1 that brains are required to compute Bayesian probabilities of states of the world, given their sense data [Worden 1995, 2024]. At level 2, there are Bayesian computational frameworks such as the Free Energy Principle [Friston Kilner & Harrison 2006; Friston 2010], defined abstractly by systems of equations; and these frameworks have neural implementations at level 3 (usually assuming McCulloch-Pitts neurons) [Parr, Pezzulo & Friston 2022].

At Marr's levels 1 and 2, the Bayesian approach has been very successful [Anderson 1990; Rao et al 2002; Krebs & Davies 1989; Kersten & Yuille 2003; Friston 2010]. So far it has not been decisively tested at the neural implementation level 3. This is because, in spite of advances in observing functioning brains using techniques such as FMRI, those data reveal changes at the level of millimetres and seconds; they do not probe the behaviour of detailed neural circuits in milliseconds. Similarly, while electrodes may probe individual or a few neurons, they do not test the behaviour of whole neural circuits. We are only now beginning to have data on large numbers of neurons at high resolution in time and space, with techniques such as two-photon calcium imaging [Grienberger et al. 2022]

The growth of static neural connectionism, following the work of Rumelhart and McLelland [1987] can be regarded as a repudiation of Marr's three-level framework; it says in effect: "Let's not bother with levels 1 and 2; let's try out some static neural nets, with simplified model neurons, and see what works.". I shall address this approach in section 8.

In this section I use the Bayesian computational framework at Marr's levels 1 and 2, to define in semi-quantitative terms what sort of computations the neural level 3 is required to do.

Any animal is required to compute Bayesian probabilities of the possible source of its sense data – to try to infer the most likely states of the world, given its sense data [Kersten & Yuille 2003]. It is required to classify the sources of its sense data – for instance to infer that 'the most likely source of that visual stimulus is an insect'; using the class 'insect' to classify an object in the world, so as to know what to do next. In order to find the most likely class 'insect' for some sense data, the brain needs to try out several other possible classes such as 'shadow', 'twig' or 'stone', to find the class with the highest Bayesian probability. In this way, the brain maintains a generative model [Friston 2010] of the hidden state of the world which causes its sense data – the most likely state of the world, in the light of Bayes' theorem.

At any moment, there is a very large number of possible states of the world. For a typical mammal, at any moment the visual field may contain, say, fifty distinct regions, each of which might contain almost any object. To identify one of these objects, the brain needs to try out, say, 3-5 possibilities in parallel. The result is that the brain needs to carry out a large number of Bayesian fits to sense data (in the broadest order of magnitude terms, 100) in parallel. This may be an underestimate. Each fit may be a complex Bayesian pattern-matching process, varying a large number of parameters to find the best fit to sense data.

This parallel Bayesian inference is hierarchical [Parr, Pezzulo & Friston 2020]. The world consists of hierarchies of parts and wholes, like tree-branch-leaf, or person-hand finger; it is often necessary to classify the parts and wholes simultaneously, by a joint bottom-up and top-down process.

Bayesian inferences are coordinated in hierarchical structures, but there is some process independence within that structure. In a typical Bayesian computational model, each inference has up to 20 inputs, including its top-down, bottom-up and possibly lateral inputs; and may involve a search for the optimal point in a space of 10 or more dimensions, depending on a complex structure of prior probabilities.

Most animals need to learn and match thousands of object patterns – for instance, to recognise the thousands of places they may visit in their habitat, to anticipate what is likely to happen in each place.

In order of magnitude terms, in each second a mammalian brain is required to do of the order of 100 partly independent Bayesian classification tasks in parallel, each one with up to 20 inputs, a large number of possible patterns to match, and many variables to be searched for the best fit. An insect brain has a somewhat reduced requirement [Strausfeld 2011]; but this is not massively reduced, for an insect like a bee; and its brain must respond, say, 5 times faster than a mammal brain. I assume that this is the kind of computation that neurons are required to do.

Qualitatively, as the results of pattern matching are hierarchical patterns, they must be represented by hierarchical data structures, of variable depth; and these hierarchical structures must be dynamically created, compared and stored.

There is a requirement at any moment of the day to rapidly retrieve from this large set of thousands of learned patterns a small subset of candidate patterns, to test them to find the best fit to current sense data. So the learned patterns may need to be stored in a data structure designed to support fast retrieval. One such data structure is a **subsumption graph**, [Zambon & Rensink 2012] a kind of association graph in which each node of the graph is itself a hierarchical pattern. This is a requirement to store a graph of graphs. Recall that this is a requirement at Marr's level 2 of logical data structures, not physical neural structures – which may implement the logical requirement in a number of different ways.

As well as storing all these data structures, there is a requirement to move them from one part of the brain to another - so as to compare one hierarchical pattern with another, to evaluate the match between the two. It is hard to see how patterns could be compared without moving them. For instance, some types of sense data have a hierarchical structure, and need to be moved from a specific sense data processing region of the brain to a multi-sensory comparison part of the brain. The bundles of axons which move information over larger distances across the brain are a prominent feature of brains.

The logical requirements defined by Marr's level 2 are a set of requirements for the neural processing level which is very challenging, both in quantitative and qualitative terms.

## 5. Neural Computational Models of the Brain

In the previous section, the requirements for computation in the brain were stated at Marr's Levels 1 and 2. They were stated in terms of Bayesian inference, particularly in the Free Energy Principle [Friston 2010].

In the Free Energy Principle and in other approaches to cognition, neural computational architectures have been proposed to meet these requirements. This section surveys some of those architectures (at a general level), to give a flavour of how they propose that the required computations are neurally implemented.

The neural computational architectures are usually illustrated by diagrams of neural circuits. These diagrams imply that the computation is done by neural message passing around those circuits. Three such neural architectures are shown in figures 3, 4, and 5.

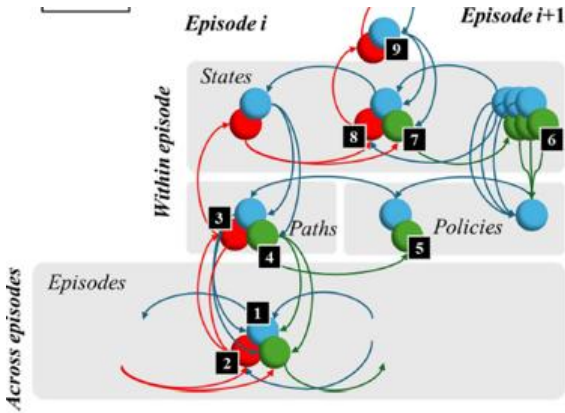

*Figure 3: Part of a neural model of episodic memory in the Free Energy Principle/Active Inference, from [Parr, Pezzulo & Friston 2020]*

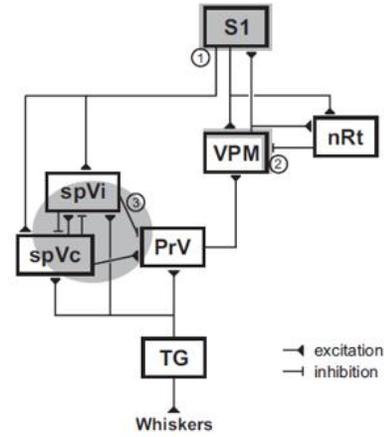

*Figure 4: A neural circuit model of vibrissal sense data in rat thalamic nuclei, from [Ego-Stengel, Le Cam & Schulz 2019]*

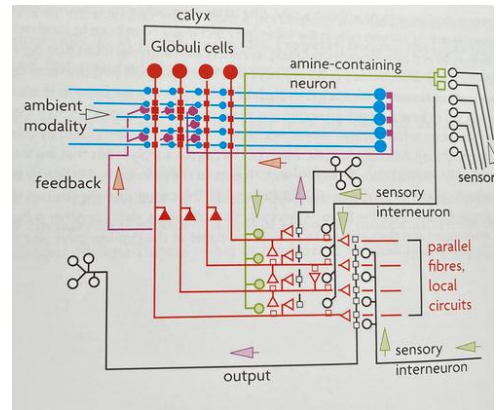

*Figure 3*

*Figure 5: A neural circuit model of olfaction in an insect mushroom body, from [Strausfeld 2012]*

There are many similar neural implementation models in the research literature. How should we interpret them? Stepping back from the details of each model, they have features in common.

The units in the diagrams typically represent individual neurons, or groups of neurons doing some simple computational function. The arrows represent links between the units, conveying information as neural spike trains. The links form chains or circuits, typically with a small number of units (2 - 5) in any chain or circuit.

Those are the elements of common accounts of computational functions, in terms of the levels of activation of each neuron or group (assumed to be a neural firing rate), and how each activation influences the activation of the next units in a chain or circuit.

Some broad assumptions of these neural models are:

- The function of each unit in a diagram is simple – capable of being done by a single McCulloch-Pitts neuron, or a small set of neurons in parallel.
- Computation is done by message-passing, and the meaning of any message is assumed to be defined statically by the type of its source unit, with some simple encoding scheme (e.g. a firing rate represents the intensity of a stimulus, or a Bayesian probability).
- The behaviour of each unit is determined by its inputs, as they are shown in the diagrams.
- Unit behaviour is determined by the dominant influence of the inputs shown in the diagrams, which is not overwhelmed by other inputs not shown.
- Alternatively: the circuits shown in the diagrams are the dominant circuits, determining the behaviour of the units. Other circuits involving the same units have less influence – providing so little background noise, that the behaviour of the circuits shown is not materially altered or degraded.

I next discuss how well these assumptions fit the neuro-anatomy of animal brains.

## 6. Numerical Parameters of Animal Brains

I shall describe some quantitative parameters of the neural structures in a brain, to ask: are they consistent with a neural architecture meeting the requirements of section 5?

This would ideally be a comparison with numerical parameters of a human or mammalian brain, because of their advanced capability. We do not yet have detailed neural connectivity data for more than a small part of the mammalian brain [Lee et al. 2025]. There is a full connectome for an adult fruit fly *Drosophila melanogaster*, from [Lin et al.2024], and all the data can be downloaded. I have written a set of Java programs to analyse those data.

The fruit fly brain has approximately 65,000 neurons and 130 million synapses; so each neuron has an average of about 2000 input synapses and 2000 output synapses. How do the synapses connect the neurons?

I first note an issue with the data [Lin et al. 2024]. While the team identified approximately 130 million synapses from electron micrographs, these were extensively proof-read to confirm which neuron was the source of each synapse, and which neuron was the target. It was only possible to unambiguously identify the source neuron for about 1 in 2 of the synapses, and to identify the target neuron for 1 in 5 synapses. Only about one in ten synapses (i.e. 12 million synapses) have both source and target neuron fully identified. I have used this subset of the synapses, and in the tables which follow I have compensated where necessary for the loss of data. I have assumed that all neurons were identified (because most neurons have many synapses, a neuron can still be identified even if most of its synapses are not identified).

I first show the distribution of input synapses per neuron, and of the number of input neurons per neuron, across all neurons in the *Drosophila* brain. These distributions are skewed, with a long tail at the high end; so it is useful to give both the mean and some percentiles, including the median (50th percentile), as in Table 1:

| Quantity | 10% ile | 50% ile (median) | 90% ile | Mean |
|---|---|---|---|---|
| Input synapses per neuron | 130 | 940 | 4,190 | 1,930 |
| Input neurons per neuron | 6 | 34 | 116 | 53 |
| Synapses per connected pair of neurons | 10 | 20 | 70 | 30 |
| Dilution factor | $1.2 \times 10^{-3}$ | $0.7 \times 10^{-2}$ | $6.5 \times 10^{-2}$ | $1.8 \times 10^{-2}$ |

*Table 1: distribution of numbers of synapses and input neurons in the drosophila brain. Numbers of synapses have been multiplied by 10 to correct for synapses whose neurons were not fully identified*

The **dilution factor** for a neuron is the ratio:

$$\frac{input\ synapses\ from\ one\ input\ neuron}{input\ synapses\ from\ all\ input\ neurons}$$

This ratio is not sensitive to any data loss. For each neuron the dilution factor measures how much the input from one of its input neurons stands out above the background noise from all other input neurons. This is important for neural computational models, where it is assumed that one or a few input neurons determine the behaviour of any neuron; if there is too much dilution, how can one or a few dominant input neurons stand out from the background?

The result of table 1 is that the largest dilution factors (the top decile) are only 6%. For any given neuron, if the input to it from one 'special' neuron is regarded as the signal, and the inputs from all other neurons are regarded as noise, then the signal: noise ratio is at best only a few percent. The signal from any special input neuron is likely to be swamped by the noise from others[2].

This result is not consistent with the neural computational models of the previous section. In those models, it is assumed that the behaviour of any neuron (or neural unit) is predominantly determined by the signal inputs from a small number of special feeder neurons or units. If the signal is only a few percent of the noise from other neurons, how can the behaviour be determined by the signal?

---

[2] Unless they are hardly firing – in which case, what are they doing?

This is a problem for conventional neural message-passing models of the brain. There are so many neural connections that any signal in a simple circuit is drowned by noise from other neurons. I shall later consider ways to rescue neural models from this problem, including various forms of parallelism. Meanwhile I explore some other consequences of the high neural connectivity of brains.

We can look downstream from any neuron, to find out how many neurons it connects to after 1, 2, 3.. synaptic links. This is to ask: if one neuron fires, and if whenever the signal arrives at any downstream neuron, that neuron also fires, how many neurons would be firing after 1, 2, 3.. synapses downstream? The answer is shown in the Table and chart of figure 6.

| Synapses | 1 | 2 | 3 | 4 | 5 |
|---|---|---|---|---|---|
| Neurons affected | 54 | 4333 | 30982 | 57671 | 62696 |

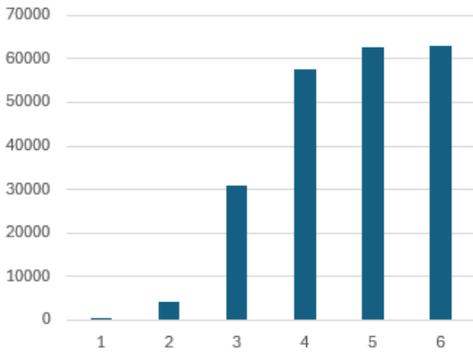

*Figure 6: Mean number of downstream neurons affected by the firing of one neuron, after following one or more synaptic links*

This shows how highly connected is the *Drosophila* brain. After only 3 synapses, approximately half the neurons in the brain are connected; after 5 synapses, nearly all 65,000 neurons are connected.

This high connectivity implies that there will be large numbers of circuits of 2, 3 and 4 neurons, of the kinds relied on in neural computational models. We can compute the distribution (across all neurons) of the numbers of 2-cycles and 3-cycles that a neuron is involved in. This is shown in Table 2:

| Quantity | 10% | 50% (median) | 90% | Mean |
|---|---|---|---|---|
| 2-cycles A-B-A per neuron A | 2 | 8 | 26 | 12 |
| Dilution product around 2-cycles | $0.7 \times 10^{-5}$ | $8 \times 10^{-5}$ | $10^{-3}$ | $2 \times 10^{-3}$ |
| 3-cycles A-B-C-A per neuron A | 5 | 130 | 865 | 300 |
| Dilution product around 3-cycles | $10^{-8}$ | $8 \times 10^{-7}$ | $5 \times 10^{-6}$ | $10^{-5}$ |

*Table 2: Properties of 2-cycles and 3-cycles in the Drosophila brain*

This shows that a typical neuron in the *Drosophila* brain is involved in an average of three hundred 3-cycles of the form A-B-C-A. The **dilution product** means that if a signal starts at neuron A, by the time it has been round the cycle A-B-C-A it has been diluted by a multiple $10^{-5}$, compared to other neural background signals into A.

With this very large number of 3-cycles, and with the extreme degree of dilution of a signal going round any 3-cycle, it is hard to argue for any functional significance of the particular 3-cycles picked out in neural computational models. How can the few 3-cycles picked out in those models be special in any way, and why is the signal going round those 3-cycles not drowned in noise? There is even more dilution for 4-cycles and higher cycles.

In summary, the large number of neuron cycles and the strong dilution factors around the cycles make it hard to argue for any special functional role of the cycles picked out in typical neural computational models.

## 7. Countering the Dilution Factor

One possible way to bridge the large gap between neural computational models and the quantitative neuroanatomy of the brain, is simple parallelism. Perhaps the units in neural computational models are not single neurons, but groups of many functionally similar neurons, each performing approximately the same function.

If two neurons are to be functionally similar, they must have similar inputs – both coming from a similar set of upstream neurons – and similar outputs, both feeding a similar set of downstream neurons.

We can make a numerical measure of this similarity, by considering the outputs of each neuron as a vector in a space of dimension 65,000 (the approximate number of neurons in the drosophila brain).

Call the two neurons being compared neuron a and neuron b. Label all other neurons with an index i, which ranges from 1 to 65,000. Neuron a connects to neuron i through $s_{ai}$ synapses. These numbers define vectors **a** and **b** in a high-dimensional space. We can define scalar products of these vectors:

$$\mathbf{a}.\mathbf{a} = \sum_i (s_{ai})^2$$

$$\mathbf{b}.\mathbf{b} = \sum_i (s_{bi})^2$$

$$\mathbf{a}.\mathbf{b} = \sum_i (s_{ai} s_{bi})$$

These scalar products are large positive integers. Define a normalized scalar product as

$$M_{ab} = \mathbf{a}.\mathbf{b} / \sqrt{(\mathbf{a}.\mathbf{a})} \sqrt{(\mathbf{b}.\mathbf{b})}$$

$M_{ab}$ has a maximum possible value of 1.0, and is a measure of the output functional similarity of neurons a and b. $M_{ab}$ is near 1.0 only if the neurons a and b have similar numbers of output synaptic connections to many other neurons. In the same way, we define an input functional similarity $N_{ab}$ for the same two neurons, and define the product $P_{ab}$ to be the overall functional similarity of neurons a and b:

$$P_{ab} = M_{ab} N_{ab}$$

The neurons a and b are functionally similar (they can act in parallel like a single unit) only if their functional similarity $P_{ab}$ is large. It is then possible to ask: for a typical neuron in the drosophila brain, how many parallel partner neurons does it have? A histogram of the number of parallel parter neurons is shown in figure 7.

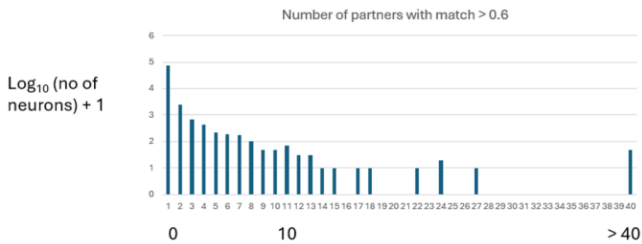

*Figure 7: Distribution of the number of similar parallel neurons of any neuron, plotted from a sample of 8000 randomly selected neurons, with functional similarity > 0.6. The graph is logarithmic, and the graph shows the logs of the numbers P of neurons with N parallel functionally similar neurons, as a function of N*

Note the logarithmic scale in this histogram. Out of a sample of 8000 neurons, about 7500 neurons have no parallel partner at all (with similarity > 0.6); 250 have just one partner; 75 have two partners, and so on. This shows that neural parallelism in the *Drosophila* brain does not solve the dilution problem. Most neurons are not in any parallel group, and only a few neurons are in small parallel groups.

(There are, however, a few groups of parallel neurons involving up to 100 neurons. These groups are the subject of a separate paper [Worden 2025a]. There are fewer than 1000 neurons out of the 65,000 involved in these large parallel groups, so they do not solve the dilution problem for most neurons)

The second solution to the problem of dilution – the solution which is advocated in this paper – is multi-state neurons. If a typical neuron has a large number of states, and if, in any one of those states, most of its input synapses are inactive, then a neuron may no longer have of the order of 50 input neurons, with a dilution factor of 2%; it could have only 2 or 3 effective input neurons, with a dilution factor close to 1. In that case, neural computational models of specific functions could work in the way that they are intended to work.

A third possible way to counter the dilution factor is to assume that actual neural connectivity is more sparse than anatomically revealed connectivity, and that most synapses in the brain are inactive simply because of Hebbian synapse changes [Hebb 1949]; to assume that Hebbian learning carves out specific neural nets from the diffuse background network of the brain.

This needs to be investigated, but it seems to have drawbacks compared to the multi-state neuron approach:

- It does not give the extra computational power of multi-state neurons (see next section).
- It is static in character: Hebbian synapse changes are typically assumed to be slow and incremental; and even if synapse strengths could be reduced or switched off rapidly, it is hard to see how they could be rapidly switched back on, without making Hebbian learning unstable.
- Static thinning of neural connections, with no chance of re-establishing connections, seems to be an inefficient use of neurons and synapses
- When trained, artificial neural nets with Hebbian learning do not go to a sparse network state.

So the preferred solution to the dilution problem is by multi-state neurons. The next section illustrates the possible computational benefits of multi-state neurons.

## 8. Bayesian Object Recognition

This section is a partly worked example of how multi-state neurons could compute more efficiently than two-state neurons, meeting a requirement which is central for all animals: to recognize the objects around them that require action (predators, food, and so on). The example is not fully worked out, and there may be ways in which the designs are not workable. They are described only to illustrate the potential of dynamic multi-state neurons compared to static neural structures using two-state neurons.

As was described in section 4, Bayesian object recognition has the following requirements:

- To recognize hierarchical structures of whole-part-subpart, such as body-limb-digit
- To integrate multi-modal sense data – vision, smell, sound, touch,…
- To learn and store large numbers of object patterns, each one as a logical tree of whole and part properties
- To compare incoming sense data in parallel against several stored patterns, choosing the best match
- To move logical tree structures from one part of the brain to another, so they can be compared
- To measure the similarity between two tree structures, to find the best matching structures

I discuss how some of these requirements might be met, first in an architecture using conventional two-state neurons, and second, using multi-state neurons. I focus on two requirements: (a) to represent data in logical tree structures by neural activity, and (b) to move tree structures from one part of the brain to another.

The first question that arises is the neural representation of a logical tree structure, with several properties on each node of the tree. Logical trees can have variable depth and branching, with any property (such as spatial extent, colour, or motion) stored on any node of the tree.

In a conventional neural architecture, we might represent a logical tree structure by a physical network of neuron connections, using some simple mapping between the two network structures. Suppose that this representation is used in two regions of the brain, called 'Left' and 'Right', and it is required to move tree structure information from the left region to the right region. This is shown in figure 8.

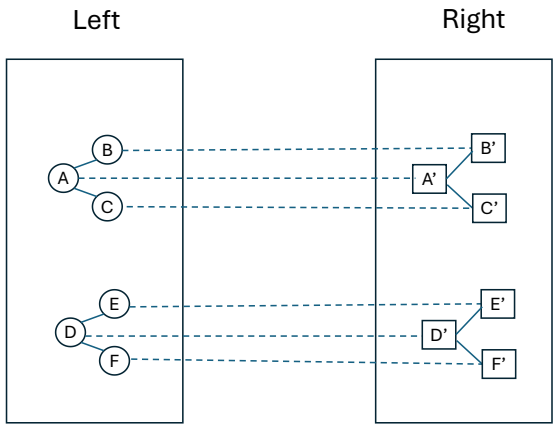

*Figure 8: Neural architecture for storing and moving logical tree information structures, using two-state neurons and physical trees which match the logical tree structure*

In both left and right regions, a logical tree structure ABC is represented by a physical tree of connected neutrons (A), (B), and (C). To convey the data values from left to right, the dashed lines represent axons from neurons (A) in the left region to their equivalents [A'] on the right. There may be several axons from (A) to [A'], conveying the values of different properties of node A (physical dimensions, colour, smell, and so on).

How many neural tree structures are needed in the left-hand region? One solution is that for every distinct tree structure, a different neural tree is required. Here, distinct tree structures may differ not only the topology of their nodes (depth and branching), but also the kinds of data stored on each node – the 'slots' such as colour intensity, size, and so on.

Each logical tree is defined by two kinds of information:

- Its **metadata**, defining the structure of the tree, and the types of information ('slots') on each node
- Its **data**, defining the values of each type of information on each node – the intensity of the colour blue, and so on.

In this model, the axons from left to right convey data, but do not convey metadata.

If a logical tree has metadata information content of M bits, then there are approximately $2^M$ different tree structures of that complexity. This means that the number of neural structures needed in each region grows exponentially with the greatest possible complexity of the trees. At any time, the great majority of these neural tree structures are not in use – which is a very inefficient use of neurons.

(An alternative is to store many copies of a maximal tree, with the maximum required depth and branching, and with all types of information stored on every node; and then using a subset of the maximal tree to store any actual required tree. This, too, is a very inefficient use of neurons).

So in this two-state neuron solution, the number of neural trees in left and right regions grows exponentially with the complexity of the objects to be recognised. Similarly, the number of axons from left to right regions grows exponentially. In neurogenesis, every axon has to be targeted precisely - from (A) to [A'] and so on. This precise targeting would be a two-state neuron solution to the neural binding problem[3].

The inefficiency of these designs derives from their choice to embody metadata (logical tree structure) physically, in the static connectivity of neurons.

As well as their great inefficiency in use of neurons, the two-state neuron designs have all the problems of signal dilution described in section 5.

I next turn to possible designs using multi-state neurons, shown in figure 9.

---

[3] It has been proposed that the binding problem [Treisman & Gelade 1980; Treisman 1998] could be solved by neural synchrony. However, neural synchrony is too weak a mechanism to give the large amount of binding information required [Shadlen and Movshon 1999].

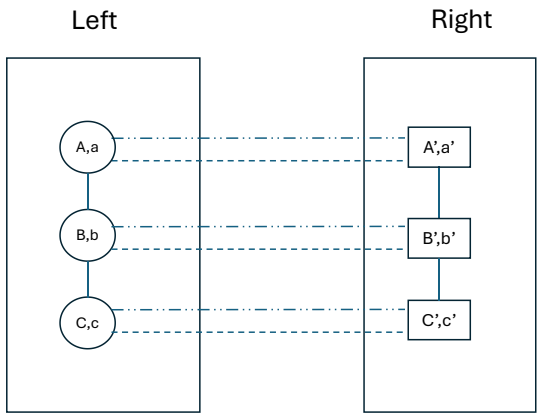

*Figure 9: Neural architecture for storing and moving logical tree information structures, using multi-state neurons and dynamic metadata*

Here, there are no special physical tree structures in the left region. The vertical lines are a shorthand, denoting that every neuron or group A, B, C… is anatomically connected to every other. Many more vertical lines should be shown for realism. Of these anatomical connections, only a small minority have active synapses at any one time, as defined by the states of the multi-state neurons. This sparse actual connectivity of the left region defines the tree structures in use.

Hence the metadata – the logical structure of the trees in use – is defined dynamically, by the states of multi-state neurons, rather than statically by anatomical connections of the neurons.

This means that a logical tree with, say, 4 nodes and 12 types of information on all the nodes could be represented by any 4 + 12 = 16 neurons in the left region. Trees can be assembled dynamically using any free neurons. This uses the multiple states of the neurons to hold metadata as follows:

- Suppose that node B of a logical data structure holds three types of information – values of three slots. That can be represented by three multi-state neurons for the data, and one or more multi-state neurons for the metadata
- Firing rates of the data neurons represent values of the data
- The metadata neurons define what the data is about – the type of each data signal.
- The group of 4 or more neurons is denoted by the circle B in the left region.
- The lower-case 'b' in the circle denotes the metadata, stored in the states of some of these multi-state neurons
- The metadata in group B denotes both the place of node B in the logical tree (that it is a child of A, and parent to D) and the types of information held on node B (blue intensity, smell,…)

To convey both data and metadata from the left region to the right, the dashed lines denote axons carrying data values, and the dot-dashed lines denote axons carrying metadata. Axons join neurons in the left and right regions in an approximately 1:1 mapping. So a group of neurons such as (B) in the left region is connected to a corresponding group of neurons [B'] in the right region.

Metadata alters the state of multi-state neurons, and that state changes more slowly than data changes. Metadata changes take place over the order of seconds or minutes, while data changes take place over tens or hundreds of milliseconds. Metadata can be conveyed between regions using a slow encoding.

This is only a partial sketch of an architecture using multi-state neurons, and some problems have not been addressed. Many tricky design problems remain to be addressed. However, it gives an indication of how the number of required neurons could scale only linearly with the number of objects that an animal can recognize in its sense data. This is more efficient than the two-state neuron architecture, where static connectivity requires a much more inefficient use of neurons. It is also a potential solution to the dilution problem.

The requirement to match large numbers of logical tree structures arises not only in general animal cognition, but also in human language. Here, the logical data structures are called **feature structures** or **constructions**, and the Bayesian maximum likelihood matching is called **unification**. The use of feature structures [Fillmore 1985, Langacker 1991, Kay 2002] and unification [Kaplan & Bresnan 1981, Gazdar et al 1985, Shieber 1986] has been studied in cognitive linguistics [Hoffman & Trousdale 2012; Worden 2025b] for many years.

## 9. Static Connectionist Architectures

The connectionist approach to modelling brains differs from the [Marr 1982] three-level approach. The connectionist approach is to ignore Marr's levels 1 and 2, and to carry out experiments at the neural implementation level, to find out what works. These experiments have typically been done using McCulloch-Pitts two-state neurons, with Hebbian learning by altering the strengths of synapses.

Recently this approach has led to spectacular progress in artificial neural nets [LeCun, Bengio & Hinton 2015], giving human-like performance in a variety of tasks, such as Large Language Models for linguistic tasks [Naveed et al. 2023]. This progress has been achieved at a cost, which includes:

- Extremely large connectionist models
- Extremely long training times, measured in 'epochs' or thousands of training examples.

- Limited human understanding of how the models perform any task, depending as it does on the weights of huge numbers of synapses, which cannot be related to performance of specific tasks.
- Unreliable performance, leading to significant numbers of 'hallucinations' and failures of common sense [Naveed et al. 2023].
- Human misunderstanding of what is happening; by Denett's [1989] Intentional Stance, people are prone to think that an AI system is 'thinking', 'feeling' and so on – generally, to erroneously attribute human-like properties to it.

Are these connectionist large networks in any sense a model for human or animal brains? Their large, diffuse networks and strong dilution are superficially similar to the connectivity seen in the *Drosophila* brain, but beyond that superficial resemblance, there is little anatomical resemblance to any animal brain. The most important reason why they are not a model for animal brains concerns learning.

Because of the very long training times, neural nets are not a model of animal learning. They require vastly more learning examples than any animal could experience in its lifetime. Whereas animals rapidly learn a few things which matter most for their fitness (how to recognize predators, food, and so on) [Anderson 1990], a typical Large Language Model learns (in some sense) everything that can be learned about a vast set of words and the statistical correlations between them, and it takes a very long time to do so.

The focused learning done by animals and people is the learning of metadata, as much as it is learning of data:

- When a bee learns about flowers as a food source, it learns that certain aspects of its sense data matter – certain wavelengths of the colour, certain properties of the visual shape, and certain components of the smell are the important variables; other variables can be ignored. Which variables matter is the metadata; the values of those variables are the data.
- When we learn to recognize a hand, we learn that some variables do not matter (such as colour; the hand might be in a glove); and other variables must be precise. There should be exactly four fingers and a thumb; nothing else will do. Which variables matter, and how precise they need to be, is metadata.
- Learning a language as humans do is largely a matter of learning metadata. Grammar is metadata about allowed sentence structure, which allows individual meanings (the meaning data of each word) to be assembled. Verbs need a subject; that is part of the metadata about any language.

Much human and animal knowledge consists of a taxonomy of types and subtypes, such as thing-animal-insect-wasp. Logically this may be represented as a subsumption graph – a directed acyclic graph, where each node of the graph is itself a tree structure describing a thing and its parts. An important component of learning is to learn the taxonomy – the structure of subsumption graph. This structure itself is metadata. Data then hangs off the graph, as slot values. The amount of metadata to be learned is comparable to the amount of data, perhaps larger.

Large neural nets learn by Hebbian changes to synapse strengths. As each synapse learns from the pulses arriving at it, it makes no distinction between data and metadata. This may be one reason why the learning is inefficient and slow; synaptic learning is affected by all data, whether or not the metadata means that it is relevant data. This may relate to the scaling laws of LLMs. Even though they do not use metadata to focus the data for specific learning, they learn a lot eventually by brute force learning.

Human reasoning is largely metadata reasoning: 'to solve this problem, I need to control these variables, and other variables do not matter'; or 'to write this program, I need to define these variables, and other variables can be left unknown'. Crucially, metadata confines the scope of the problem, and constrains the search space for solutions, by saying which variables do not matter.

Large neural nets do not explicitly use metadata to restrict the search space for solutions, as they have no concept of data versus metadata; yet somehow, they usually achieve a similar effect as a restricted-space search for solutions, by some brute force inference mechanism which nobody understands, and which still has significant error rates. Possibly the AI drawings of hands with seven fingers reflect their ignorance of metadata – such as the metadata that the number of fingers on a human hand really matters.

Large neural nets acquire both data and metadata knowledge; but the two are not distinguished, being spread across a vast number of connection strengths in a way that cannot be understood, or analysed, or controlled. It is not possible to separate the metadata knowledge and use it, as humans do.

## 10. Non-Hebbian Learning

To recap from some previous sections: The states of multi-state neurons can be used to represent metadata in the brain, such as the topology and data content of logical tree structures. Because the neuron state changes may take seconds or minutes to happen, metadata communication in animals is not suitable for the immediate response to sense data. But the timescales of metadata changes are suitable for memory and learning. If an animal can learn or remember something in a few seconds or minutes, that is useful – and is much more useful than learning in the 'epochs' required for typical neural net learning.

Hebbian learning is local and synapse-based, so it makes no distinction between data and metadata. It cannot use metadata to filter and restrict the data which it uses to learn; so it might inefficiently use all data, whereas to learn some specific thing, where only a small subset of the data is relevant. It may be that the glacial slowness of neural net learning is related to this inefficiency of Hebbian learning.

This raises the possibility that multi-state neurons might be used for a form of non-Hebbian learning in artificial neural networks. In this non-Hebbian learning process:

- The change in effective synapse strength is driven not locally by pulses arriving at the synapse, but by changes of state of a multi-state neuron fed by the synapse.
- Inputs to a multi-state neuron include both controlling synapses (which alter its state) and controlled synapses (which are switched on and off, depending on its state)
- Controlling synapses carry metadata; controlled synapses carry data
- In animal brains, state changes happen slowly, over seconds and minutes. In artificial neural networks, there is no such restriction; metadata changes could be fast.
- Learned metadata resides in the states of multi-state neurons, which may define a sparse connection topology of the network.
- There is a possibility (which has not yet been validated in a working model) of specific learning which is much faster than the Hebbian general learning of traditional artificial neural nets.
- It may be possible to inspect and understand the learned metadata, so that the neural net is no longer a black box.
- The performance of a net could be both understood and controlled by specific focused training.

I have not yet validated these possibilities in a working computational model. There may be some fundamental objection to the ideas. However, the suggestion of non-Hebbian learning is just one of many possibilities opened up by multi-state neurons in artificial neural nets; it is a promising area for investigation. There is a design space to explore.

## 11. Conclusions

In theoretical neuroscience and in artificial neural nets, neurons are usually modelled as two-state devices which make a weighted sum of their inputs to compute a binary on-off state.

As eukaryotic cells, neurons are capable of sophisticated internal computations, which are seen in morphogenesis. This paper suggests that internal neural computations continue in an adult brain, allowing each neuron in the brain to have multiple states – making state changes in the order of seconds or minutes. These state changes selectively turn off synapses, sculpting dynamic sparse neural nets which are capable of computation, from the diffuse and amorphous anatomical connections of the brain.

The use of multi-state neurons has been illustrated in Bayesian object recognition, where it supports a design with dynamic metadata, which is more efficient in its use of neurons than architectures of two-state neurons, and may solve the dilution problem.

This proposal is intended to encourage experimental neuroscientists to look for multi-state neurons in brain, and theoretical neuroscientists to explore outside the box of McCulloch-Pitts neurons.

The multi-state neuron model of the brain is contrasted with todays' artificial neural nets, which achieve impressive performance with some serious drawbacks. Two drawbacks are their very slow learning, and their black box nature; it is not possible to analyse how they do specific tasks based on their vast numbers of learned connection weights.

Artificial neural nets could be built with multi-state neurons. This might have benefits, including faster non-Hebbian learning, and a clear separation between data and metadata, leading to more understandable and controllable performance. These possibilities have not yet been demonstrated in computational models, but it is a fertile area to explore. If they could be realized, they might disrupt the economics of artificial intelligence, just as mini-computers and personal computers disrupted mainframe computing in the 1980s.